\documentclass[jpcfk,twocolumn,groupeaddress,prb]{revtex4}
\usepackage{graphicx} 

\usepackage{amsmath}

\usepackage{amsfonts}
\usepackage[version=3]{mhchem}

\usepackage{dcolumn}

\usepackage{bm}

\newcommand{\dif}{\mathrm{d}}
\newcommand{\A}{\mathrm{\AA}}

\begin{document}

\title{Hydration of Kr(aq) in Dilute and Concentrated Solutions}

\author{Mangesh I. Chaudhari}
\email{michaud@sandia.gov}
\affiliation{Center for Biological and Material Sciences, Sandia National Laboratories,  Albuquerque, NM, 87123}

\author{Dubravko Sabo}
\email{dubravko.sabo@nyu.edu}
\affiliation{Department of Chemistry, New York University, New York, NY 10003 }

\author{Lawrence R. Pratt} 
\email{lpratt@tulane.edu}
\affiliation{Department of Chemical and Biomolecular Engineering, Tulane University, New Orleans, LA, 70118}

\author{Susan B. Rempe}
\email{slrempe@sandia.gov}
\affiliation{Center for Biological and Material Sciences, Sandia National Laboratories,  Albuquerque, NM, 87185}

\date{\today}

\begin{abstract} Molecular dynamics simulations of water with both multi-Kr and
single Kr atomic solutes are carried out to implement quasi-chemical theory
evaluation of the hydration free energy of Kr(aq). This approach obtains
free energy differences reflecting Kr-Kr interactions at higher concentrations.
Those differences are negative changes in hydration free energies
with increasing concentrations at constant pressure. The changes are due to a
slight reduction of packing contributions in the higher concentration case. The
observed Kr-Kr distributions, analyzed with the extrapolation procedure of
Kr\"{u}ger, \emph{et al.},  yield a modestly attractive osmotic second virial coefficient, $B_2\approx -60~\mathrm{cm}^3$/mol. The
thermodynamic analysis interconnecting these two approaches shows that they are
closely consistent with each other, providing support for both
approaches.


\end{abstract}

\maketitle 

\section{Introduction} Here we analyze the hydration of the classic hydrophobic
solute krypton (Kr) in liquid water solution, including at elevated concentrations that
permit Kr-Kr encounters to be assessed. Kr(aq) is specifically interesting for
several reasons. Firstly, Kr  is isoelectronic with
Rb$^+$, which has been analyzed in detail.\cite{sabo2013case} Secondly,
Kr(aq) has been a target of sophisticated experimental structural
work.\cite{Filipponi:1997hj,Bowron:1998bg} Finally, Kr(aq) has been the focus of
intense theoretical attention,\cite{Watanabe:1986db,Kennan:1990dq,Pfund:2002bf,oldkr}
but the statistical mechanical theory has become much more sophisticated in
recent years.\cite{asthagiri2007non}

Here, we use molecular dynamics simulations to study Kr(aq) solutes utilizing
quasi-chemical theory. \cite{Redbook,Paulaitis:2002fd,Beck:2006wp} This theory
rigorously decomposes the free energy of hydration into packing, outer-shell and
chemical components, is independent of van der Waals
assumptions,\cite{asthagiri2007non} and thus physically clarifies the
hydrophobic solubility of Kr in water. Recent theoretical work has returned to
the basic problem of evaluating the osmotic second virial coefficients, $B_2$,
for simple hydrophobic solutes in water.\cite{Chaudhari:2013dy,koga2013osmotic}
We use the new statistical thermodynamic results here to reanalyze that
important problem.

The next section specifies the technical data for the simulations carried out here.
Succeeding sections lay out the quasi-chemical theory employed, then present
results and discussion.

\section{Computational methods}

The structural properties of interest are computed using the GROMACS molecular
dynamics simulation package.\cite{VanDerSpoel:2005hz} Simulations were carried
out in the isothermal-isobaric ($NpT$) ensemble at $T=300$ K and $p=1$ atm. The
simulation cell was a cube, and standard periodic boundary conditions were used
to mimic bulk solution.  

A single Kr atom and 16 Kr in water were simulated separately to study the
effects of solute concentration. The mole fraction of 1 Kr in 1000 water
molecules is comparable to concentrations used in experimental hydration free
energy studies; \cite{Kennan:1990dq} the higher concentration case is likely a
supersaturated solution, but that did not lead to problematic behavior. After
energy minimization and 1 ns of volume equilibration, the simulations were
extended to 20 ns of production run. Bonds involving hydrogens were fixed by the
LINCS algorithm.\cite{Hess:1997iy} The equations of motions were integrated
using a time step of 1~fs. Temperature and pressure were maintained using a
Nose-Hoover\cite{Nose:2006em} thermostat and a
Parinello-Rahman\cite{Parrinello:1981it} barostat. Interactions between water
molecules were described by the SPC/E model.\cite{Berendsen:1987uu} The Kr atoms
were treated as Lennard-Jones (LJ) spheres with parameters
$\sigma_{\mathrm{KrKr}}=3.935~\A$ and $\epsilon_{\mathrm{KrKr}}=0.4342$
kcal/mol.\cite{Jorgensen:1996vx} For KrO interactions, the mixing rules of the
OPLS force-field were followed.\cite{VanDerSpoel:2005hz} Configurations were
sampled every 0.5 ps.

\section{Theory}
The hydration free energy of Kr(aq) was computed using quasi-chemical 
theory (QCT),\cite{asthagiri2010ion,rogers2012structural}
 \begin{multline}
	\beta\mu^{\mathrm{(ex)}}_\mathrm{Kr}\left(\rho_{\mathrm{Kr}},p, T\right)  = -\ln p^{(0)}(n_{\lambda}=0) \\
	+ \ln\langle {e}^{\beta\varepsilon}\mid {n_{\lambda}=0}\rangle
	+ \ln p(n_\lambda = 0)~.
\label{eq:qct}
\end{multline} 
This equation involves packing, outer-shell and chemical contributions on
the right-hand side.  Those contributions are discussed further below. 
Eq.~\eqref{eq:qct} has been used to study hydration free
energies of water,\cite{Paliwal:2006um,Shah:2007dm,Chempath:2009dc} CH$_4$,
\cite{Asthagiri:2008uz} CF$_4$/C(CH$_3$)$_4$,\cite{asthagiri2007non},
hydrogen \cite{Sabo:2008vy} and carbon dioxide.\cite{Jiao:2011dn} We evaluate
the free energy contributions numerically by analyzing simulation trajectories.

Recognizing $p^{(0)}(n_\lambda=0)$ as the probability of finding an empty inner
shell at an arbitrary point in the fluid, the term
$-\mathrm{ln}p^{(0)}(n_{\lambda}=0) $ assesses the free energy required to form
the corresponding cavity. This packing contribution was estimated by random
placement of 30,000 spheres of radius $\lambda$, scoring the fraction of those
placements conforming to the `vacant' event.

The middle term represents interactions of outer-shell solvent molecules with a
solute in a vacant inner shell. This outer-shell contribution is approximated as
\begin{multline}
	\ln\langle {e}^{\beta\varepsilon}\mid {n_{\lambda}=0}\rangle
	\approx \beta\langle\varepsilon\mid{n_\lambda=0}\rangle \\
	+\beta^2\langle\delta\varepsilon^2\mid{n_\lambda=0\rangle}/2~.
\label{eq:gaussian}
\end{multline}
This is the 
important simplification of quasi-chemical theory
that the construction of Eq. ~\eqref{eq:qct} is designed
to achieve. With this ingredient, quasi-chemical theory is seen to be a
generalization of the van der Waals picture of
liquids.\cite{rogers2012structural,Chandler:1983up} This gaussian approximation
has been well-tested for the substantial list of applications
noted,\cite{Paliwal:2006um,Shah:2007dm,Chempath:2009dc,Asthagiri:2008uz,asthagiri2007non,Sabo:2008vy,Jiao:2011dn}
and including electrolyte solutions. \cite{zhang2013multiscale}

The right-most term of Eq.~\eqref{eq:qct} is the `chemical' contribution to the
hydration free energy. It can also be recognized as the free energy regained
with relaxation of the constraint associated with the successful formation of a
cavity at the initial step.\cite{rogers2012structural} This contribution is
estimated directly by observing Kr atoms present in the simulation. 
 
The individual contributions of Eq.~\eqref{eq:qct} depend on $\lambda$, which
sets the boundary between inner- and outer-shell solvent molecules.
The excess chemical potential, $\mu_{\mathrm{Kr}}^{\mathrm{(ex)}}$, should be independent of that boundary. Here,
we evaluate $\mu_{\mathrm{Kr}}^{\mathrm{(ex)}}$ 
for a range of $\lambda$ covering as
much structural data as statistically feasible. 

\begin{figure} 
 \begin{center}
\includegraphics[width=3.2in]{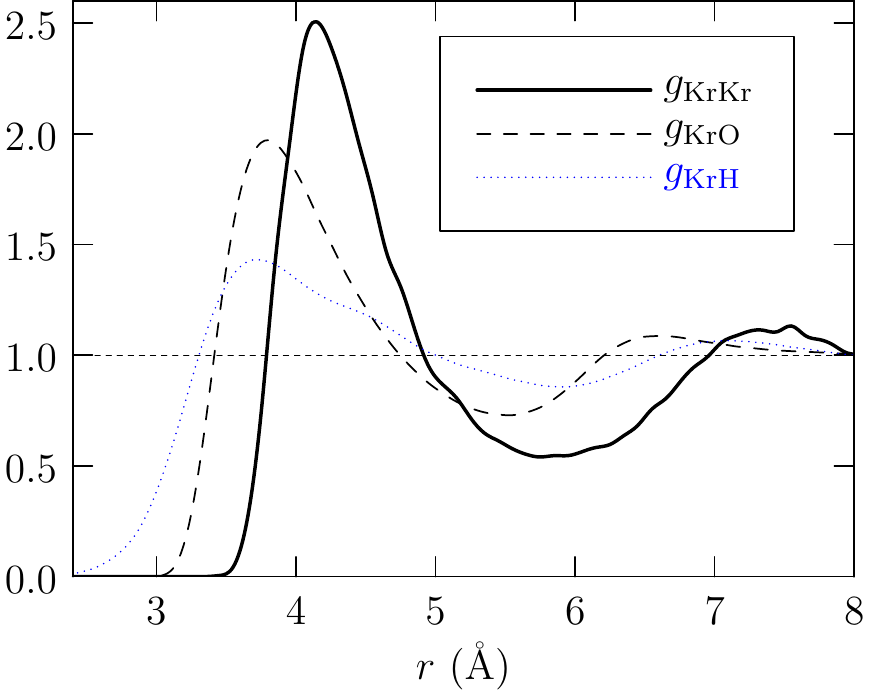}
\caption[Radial distribution function]{Radial distribution functions observed
for the aqueous Kr solution at elevated concentration, $n_{\mathrm{Kr}}=16$. The
water structure, $g_{\mathrm{OO}}\left( r \right)$ (not shown), is closely
similar to that of bulk water.\cite{}}
\label{fig:rdf}
\end{center}
\end{figure}

\begin{figure} 
 \begin{center}
\includegraphics[width=3.2in]{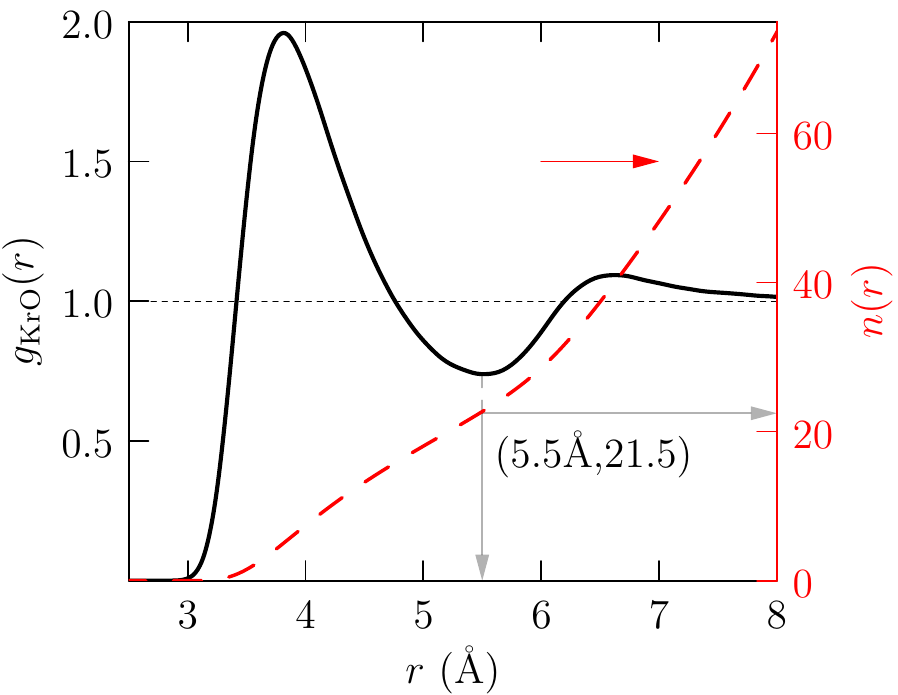}
\caption[Radial distribution function]{Higher detail for $g_{\mathrm{KrO}}(r)$
around Kr atoms (for the $n_{\mathrm{Kr}}=16$ case). The red dashed line is the
running coordination number $n(r)$, indicating an average O-occupancy of 21.5  for the
sphere within the first minimum defined by a first inner-shell radius of $\lambda=5.5$ \AA. The
principal minimum is not unusually deep, and outside the principal maximum, the
fluid is only weakly structured.}
\label{fig:running}
\end{center}
\end{figure}

\section{Results and discussion} 
\begin{figure} 
\begin{center}
\includegraphics[width=3.2in]{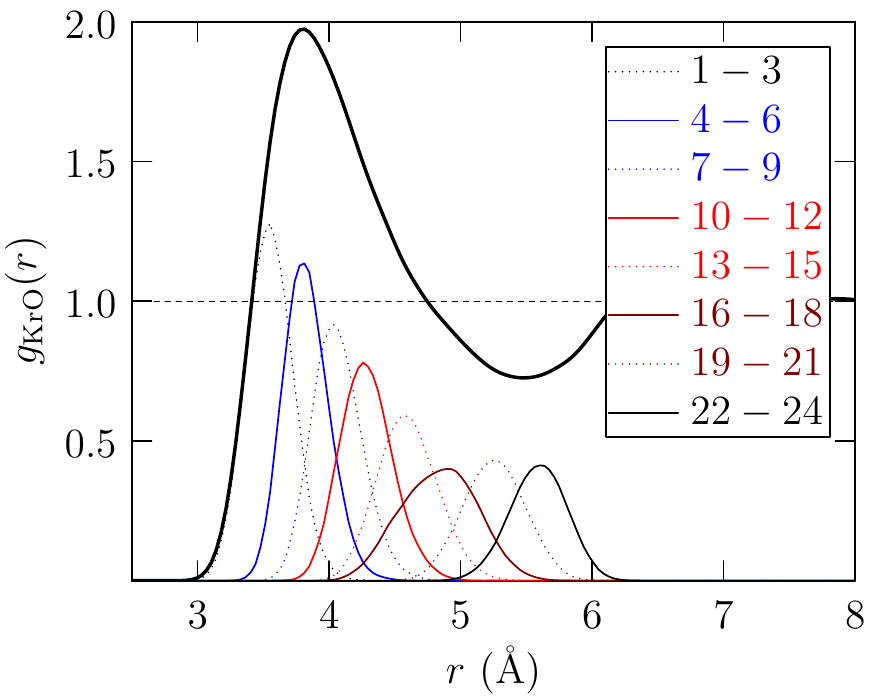}
\caption[Radial distribution function]{Neighborship decomposition of the radial
distribution $g_{\mathrm{KrO}}(r)$, from the $n_{\mathrm{Kr}}=1$ case. The
maximum is located at $r\approx 3.8$\AA. Note that the principal maximum of
$g_{\mathrm{KrO}}(r)$ is saturated by fewer than 10 nearest neighbors. See also
FIG.~\ref{fig:running}.  Outside the principal maximum, the fluid 
is only weakly structured.}
\label{fig:rdf_decomp}
\end{center}
\end{figure}

Characterizing the solution structure around
the Kr atoms, FIG. \ref{fig:rdf} shows several radial distribution functions.
Both KrO and KrH distributions display a principal maximum at about 3.7 $\A$.
The water structure, $g_{\mathrm{OO}}\left( r \right)$ (not shown), is largely
unaffected by the Kr for these calculations. A more exclusive view of the
hydration structure (FIG.~\ref{fig:running}) shows that the average O-occupancy
of the sphere within the first minimum for first inner-shell boundary of 5.5
\AA~is 21.5. The principal minimum is not unusually deep, and outside the
principal maximum the fluid is only weakly structured. Neighborship
decomposition (FIG.~\ref{fig:rdf_decomp}) shows that the principal maximum of
$g_{\mathrm{KrO}}(r)$ is saturated by fewer than 10 nearest neighbors. 

\begin{figure} 
 \begin{center}
\includegraphics[width=3.2in]{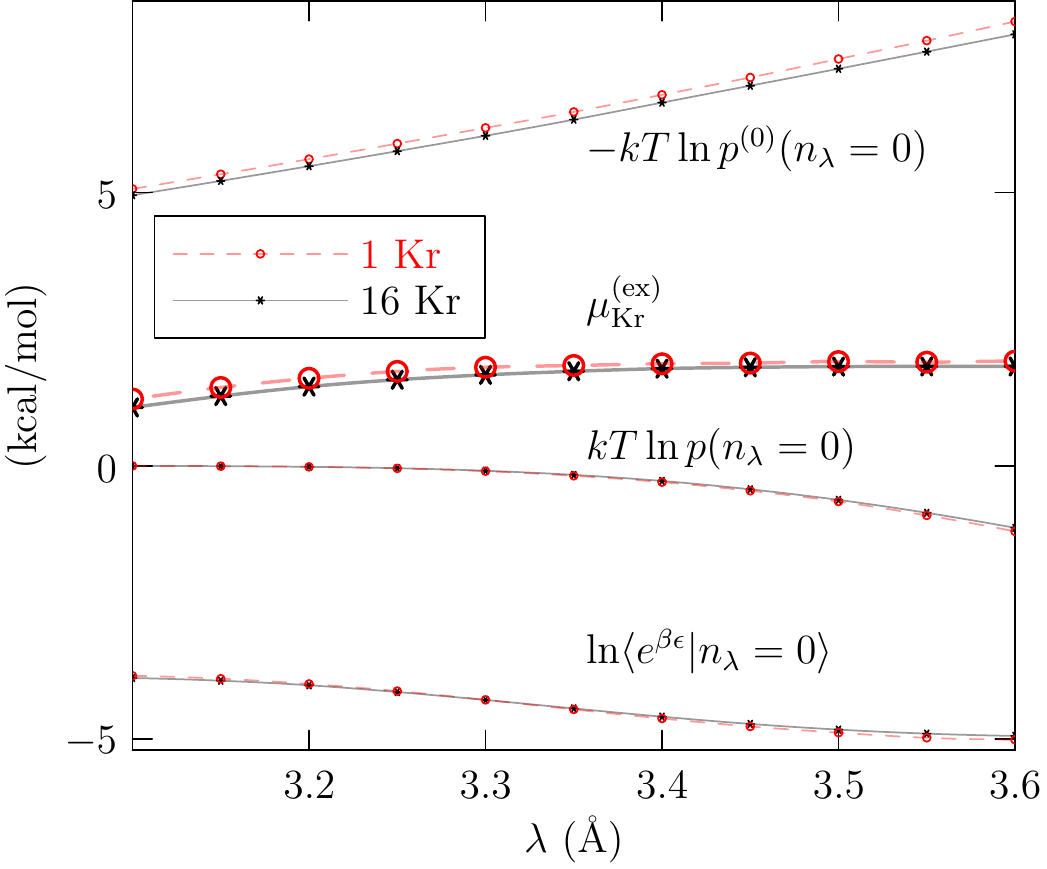}
\caption[Radial distribution function]{Packing, chemical and outer-shell
contributions to the hydration free energy of Kr were calculated numerically for
a range of $\lambda.$} 
\label{fig:pack_chem}
\end{center}
\end{figure}
\begin{figure}[h] 
 \begin{center}
\includegraphics[width=3.2in]{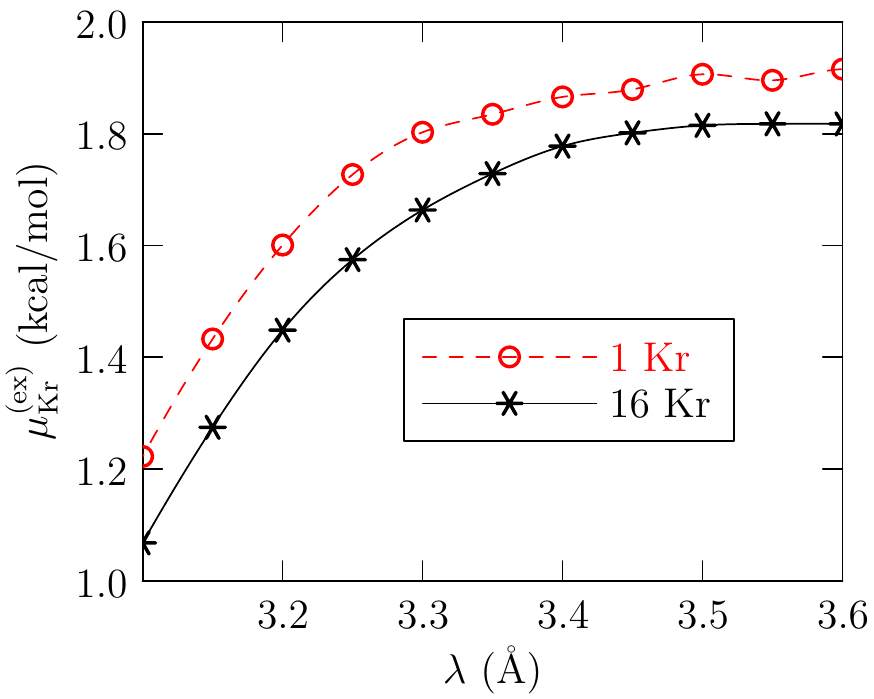}
\caption[The excess Chemical Potential]{Evaluations of hydration free energies
on the basis of quasi-chemical theory for a range of inner-shell boundaries
(3.1\AA $< \lambda <$ 3.6\AA) for the two Kr concentrations. 
$\mu_{\mathrm{Kr}}^{\mathrm{(ex)}}$ becomes insensitive to 
$\lambda$ for 3.4\AA$<\lambda<$3.6\AA .
The experimental value is 1.66 kcal/mol.\cite{young}}
\label{fig:muex}
\end{center}
\end{figure}

For solutes of van~der~Waals type generically, the natural physical idea is to
identify $\lambda$ with the van~der~Waals distance of closest
approach.\cite{asthagiri2007non,Asthagiri:2008uz} The individual contributions
(FIG.~\ref{fig:pack_chem}) lead to free energies (FIG.~\ref{fig:muex}) that are
substantially independent of $\lambda$, provided $\lambda>$3.4\AA. This
displacement is less than the position of the principal maximum of
$g_{\mathrm{KrO}}(r)$ and involves fewer than six (6) water molecules
(FIG.~\ref{fig:rdf_decomp}).

Substituting these directly calculated numerical values for the various QCT
contributions into Eq.\ref{eq:qct}, we obtain a net Kr hydration free energy of
$\mu^{ex}_{\mathrm{Kr}}=1.8-1.9$ kcal/mol, in reasonable agreement with the
experimental value of 1.66 kcal/mol.\cite{young} Results for the two
simulations, single Kr and multi-Kr, track each other. The multi-Kr results are
distinctly lower (FIG.~\ref{fig:muex}). The significant difference derives from
the packing contribution.

\begin{figure}[h] 
\begin{center}
\includegraphics[width=3.2in]{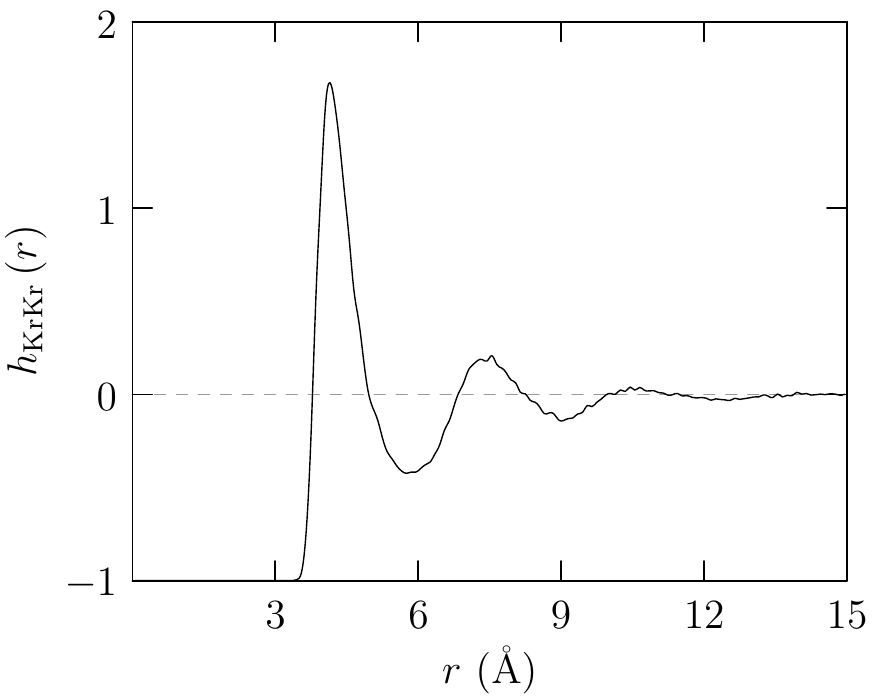}
\caption[hKrKr]{For the multi-Kr simulation, the Kr-Kr correlation function,
$h_{\mathrm{KrKr}}\left( r \right) = g_{\mathrm{KrKr}}\left( r \right) -1, $ as
in FIG.~\ref{fig:rdf}, but viewed more globally. Note the accessible $r$-range.
The second- and third-shell structures are distinct, but quantitatively 
much less structured than the results 
of Watanabe and Andersen.\cite{Watanabe:1986db}}
\label{fig:hKrKr}
\end{center}
\end{figure}

The correlation function, $h_{\mathrm{KrKr}}\left( r \right) =
g_{\mathrm{KrKr}}\left( r \right) -1$, observed in the multi-Kr simulation
(FIG.~\ref{fig:hKrKr}) provides access to the osmotic second virial coefficient,
\begin{eqnarray}
  B_2 = -\frac{1}{2}\lim_{\rho_{\mathrm{Kr}\rightarrow 0}}
  \int h_{\mathrm{KrKr}}\left( r \right) \dif^3 r ~.
\end{eqnarray}
On the basis of the McMillan-Mayer theory,\cite{zhang2013multiscale}
the integral can be identified thermodynamically as
\begin{eqnarray}
 2B_2 = \lim_{\rho_{\mathrm{Kr}\rightarrow 0}}
\left(
 \frac{\partial \beta \mu_{\mathrm{Kr}}^{\mathrm{(ex)}}}{\partial \rho_{\mathrm{Kr}}}
 \right) _{T,\mu_{\mathrm{W}}}~.
\end{eqnarray}
The partial derivative requires constancy of the water chemical potential,
$\mu_{\mathrm{W}}$, but the calculations were done at constant pressure. The
appropriate thermodynamic consideration of this distinction is 
\begin{widetext}
\begin{eqnarray}
 2B_2 = \lim_{\rho_{\mathrm{Kr}\rightarrow 0}}
 \left\lbrack
\left(
 \frac{\partial \beta \mu_{\mathrm{Kr}}^{\mathrm{(ex)}}}{\partial \rho_{\mathrm{Kr}}}
 \right) _{T,p}
 + 
 \left(
 \frac{\partial \beta \mu_{\mathrm{Kr}}^{\mathrm{(ex)}}}{\partial p}
 \right) _{T, \rho_{\mathrm{Kr}}}
  \left(
 \frac{\partial p}{\partial \rho_{\mathrm{Kr}}}
 \right) _{T, \mu_{\mathrm{W}}}\right\rbrack~.
 \label{eq:B2deriv}
\end{eqnarray}
We note that 
\begin{eqnarray}
\lim_{\rho_{\mathrm{Kr}\rightarrow 0}}   \left(
 \frac{\partial \beta p}{\partial \rho_{\mathrm{Kr}}}
 \right) _{T, \mu_{\mathrm{W}}} =
 \lim_{\rho_{\mathrm{Kr}\rightarrow 0}}   \left(
 \frac{\partial }{\partial \rho_{\mathrm{Kr}}}
 \left\lbrack 
 \beta p_\mathrm{W} \left(T, \mu_{\mathrm{W}}\right) 
 + \rho_{\mathrm{Kr}} + O\left(\rho_{\mathrm{Kr}} ^2\right)
 \right\rbrack\right) _{T, \mu_{\mathrm{W}}} 
 = 1~,
\end{eqnarray}
\end{widetext}
reflecting ideal osmotic pressure at low concentration.
Referring again to Eq.~\eqref{eq:B2deriv}, we use the notation 
\begin{eqnarray}
  v_{\mathrm{Kr}}^{\mathrm{(ex)}} =  \lim_{\rho_{\mathrm{Kr}\rightarrow 0}}\left(
 \frac{\partial \mu_{\mathrm{Kr}}^{\mathrm{(ex)}}}{\partial p}
 \right) _{T, \rho_{\mathrm{Kr}}}
\end{eqnarray}
for the \emph{excess} contribution to the partial molar volume of the solute.
Combining, we obtain
\begin{eqnarray}
 2B_2 = \lim_{\rho_{\mathrm{Kr}\rightarrow 0}}
\left(
 \frac{\partial \beta \mu_{\mathrm{Kr}}^{\mathrm{(ex)}}}{\partial \rho_{\mathrm{Kr}}}
 \right) _{T,p}
 + 
v_{\mathrm{Kr}}^{\mathrm{(ex)}}~
\label{eq:B2}
\end{eqnarray} 
in the infinite dilution limit. For water,\cite{pratt1993palma} the ideal
contribution to these partial molar volumes is about 1 cm$^3$/mol. For noble
gases in water, the partial molar volumes are positive, and
$v_{\mathrm{Kr}}^{\mathrm{(ex)}}$ is about 33 cm$^3$/mol - 1 cm$^3$/mol = 32
cm$^3$/mol.\cite{moore} The present simulation results yield a value of
$v_{\mathrm{Kr}}^{\mathrm{(ex)}} \approx$ 39 cm$^3$/mol, obtained by direct
differencing of average volumes of simulation (at constant pressure)
with/without a single Kr atom. Similarly, the leading term on the right-side of
Eq.~\eqref{eq:B2} is roughly estimated as $-160~\mathrm{cm}^3$/mol, using $\Delta \beta
\mu_{\mathrm{Kr}} \approx - 0.13$ (FIG.~\ref{fig:muex}) and the observed
$\Delta \rho_{\mathrm{Kr}} \approx 4.8\times 10^{-4}$\AA$^{-3}$.  The combined
right-side of Eq.~\eqref{eq:B2} is then about $-120~\mathrm{cm}^3$/mol.

\begin{figure}[h] 
\begin{center}
\includegraphics[width=3.2in]{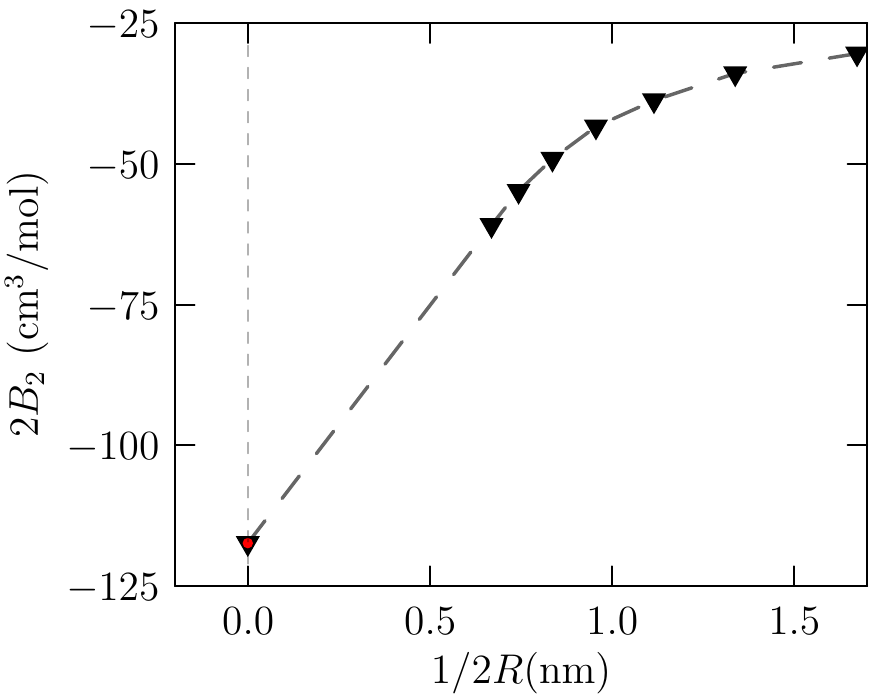}
\caption[B2extrap]{Extrapolation (Eq.~\eqref{eq:wKruger}) following Kr\"{u}ger, \emph{et
al.}\cite{kruger2012kirkwood,schnell2013apply} Computed values for $1/2R >0$
were utilized to obtain a least-squares-fit polynomial to quadratic order in
$1/2R$.  The symbol at  $1/2R =0$ is the extrapolated value and it agrees precisely
with  Eq.~\eqref{eq:B2}  to the significant figures given there.}
\label{fig:B2extrap}
\end{center}
\end{figure}

Our direct analysis of the correlation function integral
(FIG.~\ref{fig:B2extrap}) utilizes the extrapolation procedure advocated by
Kr\"{u}ger, \emph{et al.}\cite{kruger2012kirkwood,schnell2013apply} That
approach evaluates the mean and variance of Kr occupancy ($n_\mathrm{Kr}$) of a
sphere of radius $R$, which is a sub-volume of the simulation cell. That sphere
is the basis of a \emph{grand canonical} ensemble with variable $n_\mathrm{Kr}$.
Being tied to a grand canonical construction, the data analysis then corresponds
directly to the McMillan-Mayer theory theoretical
development, which proceeds through a grand ensemble.\cite{zhang2013multiscale}  The formulae 
derived on that basis 
\begin{eqnarray}
-2B_2 = \lim_{R\rightarrow \infty}
 4\pi \int_0^{2R}  h_{\mathrm{KrKr}}\left( r \right)
 w(r/2R) r^2 \dif r~,
 \label{eq:wKruger}
\end{eqnarray}
with 
\begin{eqnarray}
 w(x) = 1 - \left(\frac{3}{2}\right)x + \left(\frac{1}{2}\right)x^3~,
\end{eqnarray}
are easy to use with conventionally available distribution functions. The $1/2R
=0$ extrapolated value (FIG.~\ref{fig:B2extrap}) agrees precisely with
Eq.~\eqref{eq:B2} to the significant figures given there, the net value being
$B_2\approx -60~\mathrm{cm}^3$/mol.  In contrast, we found that direct integration 
of the distribution functions, or the $k \rightarrow 0$ extrapolation of the spatial 
Fourier transform, did not provide practical evaluations of $B_2$.

\section{Conclusion} Though Rb$^+$ is isoelectronic with Kr, the solution
structure is vastly different, and convincing inter-relation of these cases
requires serious statistical mechanical theory.\cite{sabo2013case}
Nevertheless,\cite{asthagiri2010ion} the quasi-chemical approach evaluates the
hydration free energy of Kr(aq) successfully, and in physical
terms. This approach also obtains free energy differences reflecting Kr-Kr
interactions at higher concentrations (FIG.~\ref{fig:muex}). Those differences
are negative changes in hydration free energies with
increasing concentrations at constant pressure. The changes are due to a slight
reduction of unfavorable packing contributions in the higher concentration case.
The observed Kr-Kr distributions, analyzed with the extrapolation procedure of
Kr\"{u}ger, \emph{et al.}\cite{kruger2012kirkwood,schnell2013apply}
(FIG.~\ref{fig:B2extrap}), yield $B_2\approx -60~\mathrm{cm}^3$/mol. The
thermodynamic analysis interconnecting these two approaches shows that they are
closely consistent with each other. This provides support for both approaches.

\section{Acknowledgement} We thank J. D. Weeks for telling us of Ref. 33. Sandia
is a multiprogram laboratory operated by Sandia Corporation, a Lockheed Martin
Company, for the U.S. Department of Energy's National Nuclear Security
Administration under Contract No. DE-AC04-94AL8500. The financial support of
Sandia's LDRD program and the Gulf of Mexico Research Initiative (Consortium for
Ocean Leadership Grant SA 12-05/GoMRI-002) is gratefully acknowledged.



\begin{mcitethebibliography}{34}
\providecommand*\natexlab[1]{#1}
\providecommand*\mciteSetBstSublistMode[1]{}
\providecommand*\mciteSetBstMaxWidthForm[2]{}
\providecommand*\mciteBstWouldAddEndPuncttrue
  {\def\EndOfBibitem{\unskip.}}
\providecommand*\mciteBstWouldAddEndPunctfalse
  {\let\EndOfBibitem\relax}
\providecommand*\mciteSetBstMidEndSepPunct[3]{}
\providecommand*\mciteSetBstSublistLabelBeginEnd[3]{}
\providecommand*\EndOfBibitem{}
\mciteSetBstSublistMode{f}
\mciteSetBstMaxWidthForm{subitem}{(\alph{mcitesubitemcount})}
\mciteSetBstSublistLabelBeginEnd
  {\mcitemaxwidthsubitemform\space}
  {\relax}
  {\relax}

\bibitem[Sabo \latin{et~al.}(2013)Sabo, Jiao, Varma, Pratt, and
  Rempe]{sabo2013case}
Sabo,~D., Jiao,~D., Varma,~S., Pratt,~L., and Rempe,~S. (2013) Case study of
  Rb$^+$(aq), quasi-chemical theory of ion hydration, and the no split
  occupancies rule. \emph{Ann. Rep. Section C(Physical Chemistry)} \emph{109},
  266--278\relax
\mciteBstWouldAddEndPuncttrue
\mciteSetBstMidEndSepPunct{\mcitedefaultmidpunct}
{\mcitedefaultendpunct}{\mcitedefaultseppunct}\relax
\EndOfBibitem
\bibitem[Filipponi \latin{et~al.}(1997)Filipponi, Bowron, Lobban, and
  Finney]{Filipponi:1997hj}
Filipponi,~A., Bowron,~D., Lobban,~C., and Finney,~J. (1997) {Structural
  Determination of the Hydrophobic Hydration Shell of Kr}. \emph{Phys. Rev.
  Letts.} \emph{79}, 1293--1296\relax
\mciteBstWouldAddEndPuncttrue
\mciteSetBstMidEndSepPunct{\mcitedefaultmidpunct}
{\mcitedefaultendpunct}{\mcitedefaultseppunct}\relax
\EndOfBibitem
\bibitem[Bowron \latin{et~al.}(1998)Bowron, Filipponi, Roberts, and
  Finney]{Bowron:1998bg}
Bowron,~D., Filipponi,~A., Roberts,~M., and Finney,~J. (1998) {Hydrophobic
  Hydration and the Formation of a Clathrate Hydrate}. \emph{Phys. Rev. Letts.}
  \emph{81}, 4164--4167\relax
\mciteBstWouldAddEndPuncttrue
\mciteSetBstMidEndSepPunct{\mcitedefaultmidpunct}
{\mcitedefaultendpunct}{\mcitedefaultseppunct}\relax
\EndOfBibitem
\bibitem[Watanabe and Andersen(1986)Watanabe, and Andersen]{Watanabe:1986db}
Watanabe,~K., and Andersen,~H.~C. (1986) {Molecular dynamics study of the
  hydrophobic interaction in an aqueous solution of krypton}. \emph{J. Phys.
  Chem.} \emph{90}, 795--802\relax
\mciteBstWouldAddEndPuncttrue
\mciteSetBstMidEndSepPunct{\mcitedefaultmidpunct}
{\mcitedefaultendpunct}{\mcitedefaultseppunct}\relax
\EndOfBibitem
\bibitem[Kennan and Pollack(1990)Kennan, and Pollack]{Kennan:1990dq}
Kennan,~R.~P., and Pollack,~G.~L. (1990) {Pressure dependence of the solubility
  of nitrogen, argon, krypton, and xenon in water}. \emph{J. Chem. Phys.}
  \emph{93}, 2724--2735\relax
\mciteBstWouldAddEndPuncttrue
\mciteSetBstMidEndSepPunct{\mcitedefaultmidpunct}
{\mcitedefaultendpunct}{\mcitedefaultseppunct}\relax
\EndOfBibitem
\bibitem[Pfund \latin{et~al.}(1994)Pfund, Darab, Fulton, and Ma]{Pfund:2002bf}
Pfund,~D.~M., Darab,~J.~G., Fulton,~J.~L., and Ma,~Y. (1994) {An XAFS Study of
  Strontium Ions and Krypton in Supercritical Water}. \emph{J. Phys. Chem.}
  \emph{98}, 13102--13107\relax
\mciteBstWouldAddEndPuncttrue
\mciteSetBstMidEndSepPunct{\mcitedefaultmidpunct}
{\mcitedefaultendpunct}{\mcitedefaultseppunct}\relax
\EndOfBibitem
\bibitem[Ashbaugh \latin{et~al.}(2003)Ashbaugh, Asthagiri, Pratt, and
  Rempe]{oldkr}
Ashbaugh,~H.~S., Asthagiri,~D., Pratt,~L.~R., and Rempe,~S.~B. (2003)
  {Hydration of krypton and consideration of clathrate models of hydrophobic
  effects from the perspective of quasi-chemical theory.} \emph{Biophys. Chem.}
  \emph{105}, 323--338\relax
\mciteBstWouldAddEndPuncttrue
\mciteSetBstMidEndSepPunct{\mcitedefaultmidpunct}
{\mcitedefaultendpunct}{\mcitedefaultseppunct}\relax
\EndOfBibitem
\bibitem[Asthagiri \latin{et~al.}(2007)Asthagiri, Ashbaugh, Piryatinski,
  Paulaitis, and Pratt]{asthagiri2007non}
Asthagiri,~D., Ashbaugh,~H., Piryatinski,~A., Paulaitis,~M., and Pratt,~L.
  (2007) Non-van der Waals treatment of the hydrophobic solubilities of CF$_4$.
  \emph{J. Am. Chem. Soc.} \emph{129}, 10133--10140\relax
\mciteBstWouldAddEndPuncttrue
\mciteSetBstMidEndSepPunct{\mcitedefaultmidpunct}
{\mcitedefaultendpunct}{\mcitedefaultseppunct}\relax
\EndOfBibitem
\bibitem[Pratt and Rempe(1999)Pratt, and Rempe]{Redbook}
Pratt,~L.~R., and Rempe,~S.~B. (1999) Quasi-chemical theory and implicit
  solvent models for simulations. \emph{AIP Conference Proceedings} \emph{492},
  172--201\relax
\mciteBstWouldAddEndPuncttrue
\mciteSetBstMidEndSepPunct{\mcitedefaultmidpunct}
{\mcitedefaultendpunct}{\mcitedefaultseppunct}\relax
\EndOfBibitem
\bibitem[Paulaitis and Pratt(2002)Paulaitis, and Pratt]{Paulaitis:2002fd}
Paulaitis,~M.~E., and Pratt,~L. (2002) {Hydration theory for molecular
  biophysics}. \emph{Adv. Prot. Chem.} \emph{62}, 283--310\relax
\mciteBstWouldAddEndPuncttrue
\mciteSetBstMidEndSepPunct{\mcitedefaultmidpunct}
{\mcitedefaultendpunct}{\mcitedefaultseppunct}\relax
\EndOfBibitem
\bibitem[Beck \latin{et~al.}(2006)Beck, Paulaitis, and Pratt]{Beck:2006wp}
Beck,~T.~L., Paulaitis,~M.~E., and Pratt,~L.~R. \emph{{The Potential
  Distribution Theorem and Models of Molecular Solutions}}; Cambridge
  University Press, 2006\relax
\mciteBstWouldAddEndPuncttrue
\mciteSetBstMidEndSepPunct{\mcitedefaultmidpunct}
{\mcitedefaultendpunct}{\mcitedefaultseppunct}\relax
\EndOfBibitem
\bibitem[Chaudhari \latin{et~al.}(2013)Chaudhari, Holleran, Ashbaugh, and
  Pratt]{Chaudhari:2013dy}
Chaudhari,~M.~I., Holleran,~S.~A., Ashbaugh,~H.~S., and Pratt,~L.~R. (2013)
  {Molecular-scale hydrophobic interactions between hard-sphere reference
  solutes are attractive and endothermic}. \emph{Proc. Natl. Acad. Sci. USA}
  \emph{110}, 20557--20562\relax
\mciteBstWouldAddEndPuncttrue
\mciteSetBstMidEndSepPunct{\mcitedefaultmidpunct}
{\mcitedefaultendpunct}{\mcitedefaultseppunct}\relax
\EndOfBibitem
\bibitem[Koga(2013)]{koga2013osmotic}
Koga,~K. (2013) Osmotic second virial coefficient of methane in water. \emph{J.
  Phys. Chem. B} \emph{117}, 12619--12624\relax
\mciteBstWouldAddEndPuncttrue
\mciteSetBstMidEndSepPunct{\mcitedefaultmidpunct}
{\mcitedefaultendpunct}{\mcitedefaultseppunct}\relax
\EndOfBibitem
\bibitem[Van Der~Spoel \latin{et~al.}(2005)Van Der~Spoel, Lindahl, Hess,
  Groenhof, Mark, and Berendsen]{VanDerSpoel:2005hz}
Van Der~Spoel,~D., Lindahl,~E., Hess,~B., Groenhof,~G., Mark,~A.~E., and
  Berendsen,~H. J.~C. (2005) {GROMACS: Fast, flexible, and free}. \emph{J.
  Comp. Chem.} \emph{26}, 1701--1718\relax
\mciteBstWouldAddEndPuncttrue
\mciteSetBstMidEndSepPunct{\mcitedefaultmidpunct}
{\mcitedefaultendpunct}{\mcitedefaultseppunct}\relax
\EndOfBibitem
\bibitem[Hess \latin{et~al.}(1997)Hess, Bekker, and Berendsen]{Hess:1997iy}
Hess,~B., Bekker,~H., and Berendsen,~H. (1997) {LINCS: a linear constraint
  solver for molecular simulations}. \emph{J. Comp. Chem.} \emph{8},
  1463--1472\relax
\mciteBstWouldAddEndPuncttrue
\mciteSetBstMidEndSepPunct{\mcitedefaultmidpunct}
{\mcitedefaultendpunct}{\mcitedefaultseppunct}\relax
\EndOfBibitem
\bibitem[Nos{\'e}(1984)]{Nose:2006em}
Nos{\'e},~S. (1984) {A molecular dynamics method for simulations in the
  canonical ensemble}. \emph{Mol. Phys.} \emph{52}, 255--268\relax
\mciteBstWouldAddEndPuncttrue
\mciteSetBstMidEndSepPunct{\mcitedefaultmidpunct}
{\mcitedefaultendpunct}{\mcitedefaultseppunct}\relax
\EndOfBibitem
\bibitem[Parrinello and Rahman(1981)Parrinello, and Rahman]{Parrinello:1981it}
Parrinello,~M., and Rahman,~A. (1981) {Polymorphic transitions in single
  crystals: A new molecular dynamics method}. \emph{J. Appl. Phys.} \emph{52},
  7182--7190\relax
\mciteBstWouldAddEndPuncttrue
\mciteSetBstMidEndSepPunct{\mcitedefaultmidpunct}
{\mcitedefaultendpunct}{\mcitedefaultseppunct}\relax
\EndOfBibitem
\bibitem[Berendsen and Grigera(1987)Berendsen, and Grigera]{Berendsen:1987uu}
Berendsen,~H., and Grigera,~J.~R. (1987) {The missing term in effective pair
  potentials}. \emph{J. Phys. Chem.} \emph{91}, 6269--6271\relax
\mciteBstWouldAddEndPuncttrue
\mciteSetBstMidEndSepPunct{\mcitedefaultmidpunct}
{\mcitedefaultendpunct}{\mcitedefaultseppunct}\relax
\EndOfBibitem
\bibitem[Jorgensen and Maxwell(1996)Jorgensen, and Maxwell]{Jorgensen:1996vx}
Jorgensen,~W.~L., and Maxwell,~D.~S. (1996) {Development and testing of the
  OPLS all-atom force field on conformational energetics and properties of
  organic liquids}. \emph{J. Am. Chem. Soc.} \emph{118}, 11225--11236\relax
\mciteBstWouldAddEndPuncttrue
\mciteSetBstMidEndSepPunct{\mcitedefaultmidpunct}
{\mcitedefaultendpunct}{\mcitedefaultseppunct}\relax
\EndOfBibitem
\bibitem[Asthagiri \latin{et~al.}(2010)Asthagiri, Dixit, Merchant, Paulaitis,
  Pratt, Rempe, and Varma]{asthagiri2010ion}
Asthagiri,~D., Dixit,~P., Merchant,~S., Paulaitis,~M., Pratt,~L., Rempe,~S.,
  and Varma,~S. (2010) Ion selectivity from local configurations of ligands in
  solutions and ion channels. \emph{Chem. Phys. Letts.} \emph{485}, 1--7\relax
\mciteBstWouldAddEndPuncttrue
\mciteSetBstMidEndSepPunct{\mcitedefaultmidpunct}
{\mcitedefaultendpunct}{\mcitedefaultseppunct}\relax
\EndOfBibitem
\bibitem[Rogers \latin{et~al.}(2012)Rogers, Jiao, Pratt, and
  Rempe]{rogers2012structural}
Rogers,~D.~M., Jiao,~D., Pratt,~L., and Rempe,~S.~B. (2012) Structural models
  and molecular thermodynamics of hydration of ions and small molecules.
  \emph{Annu. Rep. Comput. Chem.} \emph{8}, 71--128\relax
\mciteBstWouldAddEndPuncttrue
\mciteSetBstMidEndSepPunct{\mcitedefaultmidpunct}
{\mcitedefaultendpunct}{\mcitedefaultseppunct}\relax
\EndOfBibitem
\bibitem[Paliwal \latin{et~al.}(2006)Paliwal, Asthagiri, Pratt, Ashbaugh, and
  Paulaitis]{Paliwal:2006um}
Paliwal,~A., Asthagiri,~D., Pratt,~L.~R., Ashbaugh,~H.~S., and Paulaitis,~M.~E.
  (2006) {An analysis of molecular packing and chemical association in liquid
  water using quasichemical theory}. \emph{J. Chem. Phys.} \emph{124},
  224502\relax
\mciteBstWouldAddEndPuncttrue
\mciteSetBstMidEndSepPunct{\mcitedefaultmidpunct}
{\mcitedefaultendpunct}{\mcitedefaultseppunct}\relax
\EndOfBibitem
\bibitem[Shah \latin{et~al.}(2007)Shah, Asthagiri, Pratt, and
  Paulaitis]{Shah:2007dm}
Shah,~J.~K., Asthagiri,~D., Pratt,~L.~R., and Paulaitis,~M.~E. (2007)
  {Balancing local order and long-ranged interactions in the molecular theory
  of liquid water}. \emph{J. Chem. Phys.} \emph{127}, 144508\relax
\mciteBstWouldAddEndPuncttrue
\mciteSetBstMidEndSepPunct{\mcitedefaultmidpunct}
{\mcitedefaultendpunct}{\mcitedefaultseppunct}\relax
\EndOfBibitem
\bibitem[Chempath and Pratt(2009)Chempath, and Pratt]{Chempath:2009dc}
Chempath,~S., and Pratt,~L.~R. (2009) Distribution of Binding Energies of a
  Water Molecule in the Water Liquid-Vapor Interface. \emph{J. Phys. Chem. B}
  \emph{113}, 4147--4151\relax
\mciteBstWouldAddEndPuncttrue
\mciteSetBstMidEndSepPunct{\mcitedefaultmidpunct}
{\mcitedefaultendpunct}{\mcitedefaultseppunct}\relax
\EndOfBibitem
\bibitem[Asthagiri \latin{et~al.}(2008)Asthagiri, Merchant, and
  Pratt]{Asthagiri:2008uz}
Asthagiri,~D., Merchant,~S., and Pratt,~L.~R. (2008) {Role of attractive
  methane-water interactions in the potential of mean force between methane
  molecules in water}. \emph{J. Chem. Phys.} \emph{128}, 244512\relax
\mciteBstWouldAddEndPuncttrue
\mciteSetBstMidEndSepPunct{\mcitedefaultmidpunct}
{\mcitedefaultendpunct}{\mcitedefaultseppunct}\relax
\EndOfBibitem
\bibitem[Sabo \latin{et~al.}(2008)Sabo, Varma, Martin, and Rempe]{Sabo:2008vy}
Sabo,~D., Varma,~S., Martin,~M.~G., and Rempe,~S.~B. (2008) {Studies of the
  thermodynamic properties of hydrogen gas in bulk water}. \emph{J. Phys. Chem.
  B} \emph{112}, 867—876\relax
\mciteBstWouldAddEndPuncttrue
\mciteSetBstMidEndSepPunct{\mcitedefaultmidpunct}
{\mcitedefaultendpunct}{\mcitedefaultseppunct}\relax
\EndOfBibitem
\bibitem[Jiao and Rempe(2011)Jiao, and Rempe]{Jiao:2011dn}
Jiao,~D., and Rempe,~S.~B. (2011) {CO$_2$ solvation free energy using
  quasi-chemical theory}. \emph{J. Chem. Phys.} \emph{134}, 224506\relax
\mciteBstWouldAddEndPuncttrue
\mciteSetBstMidEndSepPunct{\mcitedefaultmidpunct}
{\mcitedefaultendpunct}{\mcitedefaultseppunct}\relax
\EndOfBibitem
\bibitem[Chandler \latin{et~al.}(1983)Chandler, Weeks, and
  Andersen]{Chandler:1983up}
Chandler,~D., Weeks,~J.~D., and Andersen,~H. (1983) {Van der~Waals Picture of
  Liquids, Solids, and Phase-Transformations}. \emph{Science} \emph{220},
  787--794\relax
\mciteBstWouldAddEndPuncttrue
\mciteSetBstMidEndSepPunct{\mcitedefaultmidpunct}
{\mcitedefaultendpunct}{\mcitedefaultseppunct}\relax
\EndOfBibitem
\bibitem[Zhang \latin{et~al.}(2013)Zhang, You, and Pratt]{zhang2013multiscale}
Zhang,~W., You,~X., and Pratt,~L.~R. (2013) Multiscale Theory in the Molecular
  Simulation of Electrolyte Solutions. \emph{J. Phys. Chem. B} \emph{118},
  7730--7738\relax
\mciteBstWouldAddEndPuncttrue
\mciteSetBstMidEndSepPunct{\mcitedefaultmidpunct}
{\mcitedefaultendpunct}{\mcitedefaultseppunct}\relax
\EndOfBibitem
\bibitem[Young(1981)]{young}
Young,~C. \emph{Hydrogen and deuterium (Solubility data series)}; Pergamon
  Press, 1981\relax
\mciteBstWouldAddEndPuncttrue
\mciteSetBstMidEndSepPunct{\mcitedefaultmidpunct}
{\mcitedefaultendpunct}{\mcitedefaultseppunct}\relax
\EndOfBibitem
\bibitem[Pratt and Pohorille(1993)Pratt, and Pohorille]{pratt1993palma}
Pratt,~L., and Pohorille,~A. In \emph{Proceedings of the EBSA 1992
  International Workshop on Water-Biomolecule Interactions}; Palma,~M.~U.,
  Palma-Vittorelli,~M.~B., and Parak,~F., Eds.; Societ{\'a} Italiana de Fisica,
  Bologna, 1993; pp 261--268\relax
\mciteBstWouldAddEndPuncttrue
\mciteSetBstMidEndSepPunct{\mcitedefaultmidpunct}
{\mcitedefaultendpunct}{\mcitedefaultseppunct}\relax
\EndOfBibitem
\bibitem[Moore \latin{et~al.}(1982)Moore, Battino, Handa, and Wilhelm]{moore}
Moore,~J., Battino,~T.~R., Handa,~Y.~P., and Wilhelm,~E. (1982) \emph{J. Chem.
  Eng. Data} \emph{27}\relax
\mciteBstWouldAddEndPuncttrue
\mciteSetBstMidEndSepPunct{\mcitedefaultmidpunct}
{\mcitedefaultendpunct}{\mcitedefaultseppunct}\relax
\EndOfBibitem
\bibitem[Kr\"{u}̈ger \latin{et~al.}(2012)Kr\"{u}̈ger, Schnell, Bedeaux,
  Kjelstrup, Vlugt, and Simon]{kruger2012kirkwood}
Kr\"{u}̈ger,~P., Schnell,~S.~K., Bedeaux,~D., Kjelstrup,~S., Vlugt,~T.~J., and
  Simon,~J.-M. (2012) Kirkwood--Buff Integrals for Finite Volumes. \emph{J.
  Phys. Chem. Letts.} \emph{4}, 235--238\relax
\mciteBstWouldAddEndPuncttrue
\mciteSetBstMidEndSepPunct{\mcitedefaultmidpunct}
{\mcitedefaultendpunct}{\mcitedefaultseppunct}\relax
\EndOfBibitem
\bibitem[Schnell \latin{et~al.}(2013)Schnell, Englebienne, Simon, Kr{\"u}ger,
  Balaji, Kjelstrup, Bedeaux, Bardow, and Vlugt]{schnell2013apply}
Schnell,~S.~K., Englebienne,~P., Simon,~J.-M., Kr{\"u}ger,~P., Balaji,~S.~P.,
  Kjelstrup,~S., Bedeaux,~D., Bardow,~A., and Vlugt,~T.~J. (2013) How to apply
  the Kirkwood--Buff theory to individual species in salt solutions.
  \emph{Chem. Phys. Letts.} \emph{582}, 154--157\relax
\mciteBstWouldAddEndPuncttrue
\mciteSetBstMidEndSepPunct{\mcitedefaultmidpunct}
{\mcitedefaultendpunct}{\mcitedefaultseppunct}\relax
\EndOfBibitem
\end{mcitethebibliography}

\providecommand{\latin}[1]{#1}
\providecommand*\mcitethebibliography{\thebibliography}
\csname @ifundefined\endcsname{endmcitethebibliography}
  {\let\endmcitethebibliography\endthebibliography}{}

\end{document}